# Discrete element model for high strain rate deformations of snow

T. Theile[1], D. Szabo, C. Willibald, M. Schneebeli

*WSL Institute for Snow and Avalanche Research SLF, Flüelastrasse 11, Davos Dorf, Switzerland,*

*thiemotheile@gmx.de*

## Abstract

In engineering applications snow often undergoes large and fast deformations. During these deformations the snow transforms from a sintered porous material into a granular material. In order to capture the fundamental mechanical behavior of this process a discrete element (DE) model is the physically most appropriate. It explicitly includes all the relevant components: the snow microstructure, consisting of bonded grains, the breaking of the bonds and the following rearrangement and interaction of the loose grains. We developed and calibrated a DE snow model based on the open source DE code liggghts. In the model snow grains are represented by randomly distributed elastic spheres connected by elastic-brittle bonds. This bonded structure corresponds to sintered snow. After applying external forces, the stresses in the bonds might exceed their strength, the bonds break, and we obtain loose particles, corresponding to granular snow. Model parameters can be divided into temperature dependent material parameters and snow type dependent microstructure parameters. Material parameters are elastic properties of the particles and bonds, coefficient of friction and coefficient of restitution of the particles and strength of the bonds. Microstructure parameters are density of the initial packing, rolling friction of the particles and diameter of the bonds. The model was calibrated by angle of repose experiments and several high strain rate mechanical tests, performed in a cold laboratory. We demonstrate the performance of the DE snow model by the simulation of a combined compression and shear deformation of different snow types with large strains. The model successfully reproduces the experiments. Most characteristics of the mechanical snow behavior are captured by the model, like the fracture behavior, the differences between low and high density snow, the granular shear flow or the densification of low density snow. The model is promising to simulate arbitrary high strain rate

1 Corresponding author. E-mail address: thiemotheile@gmx.de

processes for a wide range of snow types, and thus seems useful to be applied to different snow engineering problems.

## Introduction

A model, describing the mechanical behavior of snow, is useful for many snow related engineering problems. From snow removal equipment such as snow plows or snow blowers, constructions made of snow to mobility on snow with winter tires or chain drives. The prediction of reaction forces, deformation and failure of snow can help optimizing such snow related products. However, a general mechanical snow model does not exist. Mainly three characteristics of snow make its modelling a difficult task: the wide range of different snow types, the complex rheology of ice and under high stresses the transformation from a sintered solid material to a granular material.

Snow on the ground consists of irregularly shaped, bonded ice crystals which form a complex three dimensional continuous microstructure. The wide range of different snow types follows from the wide range of different microstructures. The size of the single ice crystal varies from a tenth of a millimeter to several millimeters, the shape varies from rounded to faceted and even hollow cup-like crystals. The bonds between the single crystals grow over time and vary from unbonded to strongly sintered. And the relative density varies from 5% for fresh snow to 60% for compacted snow. This wide range of snow types and microstructures needs to be considered in a mechanical snow model somehow. Beside the microstructure also the rheology of ice determines the mechanical behavior of snow. Ice is an elasto-viscoplastic material. Even under small stresses ice continuously creeps. Especially this creep behavior is difficult to model. It is nonlinear, highly anisotropic (on the crystal scale) and temperature dependent. In contrast, the elastic behavior is linear (Hooke's law) and shows only slight anisotropy and temperature dependence. Elastic deformation dominates only for fast deformations, for strain rates faster than $10^{-3}$ 1/s (Narita, 1983). Under high stresses bonds between the single ice crystals break and the snow transforms from a sintered solid material to a granular material going along with a huge change in its mechanical behavior. While the rheology of sintered snow is directly linked to its microstructure and the rheology of ice, the mechanical behavior of granular snow follows mainly from the interaction and friction of the loose ice grains.

A general model including all these characteristics of snow would be limited to very small simulation volumes or require enormous computing power. Therefore reasonable simplifications of the model are needed to obtain an applicable model. The model presented in this paper is intended to be applied on engineering problems like mentioned above. These problems have in common that the snow undergoes large deformations under high strain rates. Thereby the deformation and failure of the sintered snow and the behavior of the granular snow both have to be considered. Most existing models are limited to the description of the sintered snow. Constitutive equations (Shapiro et al., 1997) or finite element models (Hagenmuller et al., 2014) (Theile et al., 2011) are not well suited to describe both the sintered and granular state of snow. For this purpose DE models are predestined. DE models predict the macroscopic behavior of granular materials based on microscopic particle interactions. By adding elastic-brittle bonds between the particles also the sintered state of granular materials can be modeled (Potyondy and Cundall, 2004). The biggest challenge in modelling snow by DEs is how to consider the complex microstructure of snow.

Up to now four DE snow models have been developed. The model presented by Johnson and Hopkins (Johnson and Hopkins, 2005) included the creation and breaking of bonds and the creep behavior of ice. The model was applied to simulate the settlement of snow. However, it was never calibrated to high strain rate deformations. Michael et al. (Michael, 2014) developed a powerful DE snow model including many characteristics of snow. However, the complexity of the material model prevents applying the model to large engineering problems. Furthermore different load cases had to be treated with different parameter sets. Another snow model was developed by Hagenmuller et al. (Hagenmuller et al., 2015), focusing on an exact representation of the microstructure. The microstructure of real snow samples was imaged by micro tomography and then approximated by spheres with high accuracy. The focus of this model is to gain a better understanding of how the microstructure influences the mechanical behavior of snow. Due to the exactness of the microstructure representation it is limited to snow volumes of only a few cubic-millimeters. In contrast the model presented by Gaume et al. (Gaume et al., 2015) aims for simulating the snowpack on the scale of meters. The model is two dimensional and reproduces the crack propagation in weak snowpack layers successfully. The model does not consider the microstructure of snow but shows the potential of DE snow modelling on a larger scale. None of the existing models is capable to consider the high strain rate brittle behavior of snow on a scale that is relevant to engineering applications with the relevant micromechanical processes. Compared to the existing DE models, the model presented in this paper has a broader validation. With one set of parameters a wide range of snow types and load cases can successfully be modeled. The microstructure representation is less accurate than in the model presented by Hagenmuller et al. (Hagenmuller et al., 2015) with the benefit of faster computing times, but more accurate than in the model presented by Gaume et al. (Gaume et al., 2015).

We have chosen a bonded DE model to describe both the solid and granular state of snow. We use the open source DE code liggghts (Kloss et al., 2012). Our model contains strong simplifications regarding the material model of ice and the approximation of the microstructure. We limit the model to high strain rate deformations where creep deformation can be neglected. The particles in our model can deform purely elastically and the bonds are modeled as elastic brittle beams. The formation of new bonds is not considered in our model. The microstructure is also simplified [Fig. 1]. Each snow grain is represented by a sphere in the model. However, some important characteristics of the microstructure are considered in our model, like density, size of the bonds and shape of the snow grains by the rolling friction parameter.

The central question of this work is how well the model can reproduce snow behavior under a wide range of conditions, especially with respect to the question how a “simple” sphere system can reproduce the complex microstructure of snow. Abstract model parameters are used to imitate characteristics of the complex microstructure, like the rolling friction parameter which considers the non-sphericity of the real snow grains. These parameters were calibrated by different experiments. Angle of repose experiments were conducted to analyze the granular behavior of snow, allowing us to estimate the rolling friction and ice-ice friction in our model. Furthermore we conducted compression experiments with differently sintered snow to estimate the bond size parameter. For a broad validation of the model we performed direct shear tests with different snow types. During the direct shear test the snow undergoes compression and shear deformation with large strains of up to 80%, where the snow changes from sintered to granular.

The strength of the model is that the same set of parameters can be used for a wide variety of snow types and load cases with large strains. The reason for this is that the basic physics are captured by the model: the bonded behavior, the breaking of the bonds and the granular behavior. The snow microstructure is simplified a lot in the model, allowing fast simulations of relatively large volumes. Still different snow types can be considered by adjusting the initial density, the bond size and the rolling friction. The model is available by downloading the simulation tool from github (https://github.com/richti83/LIGGGHTS-WITH-BONDS) and with the input scripts of the most important models which can be found in the supplementary material.

# Methods

## Model

In this section the most important ingredients of the snow model are presented: the simulation tool we used, the material model and the material parameters, the microstructure approximation and the boundary conditions.

## Simulation tool and DE model

As a simulation tool we used the open source DE model liggghts (Kloss et al., 2012) with a bond-extension [retrieved from https://github.com/richti83/LIGGGHTS-WITH-BONDS]. The bond-extension is based on a publication by Potyondy and Cundall (Potyondy and Cundall, 2004). We used an elastic-brittle material model. The snow grains are represented by monodisperse elastic spheres connected by elastic-brittle bonds. Viscous deformations are not included. This limits our model to high strain rate deformations of snow, where creep behavior can be neglected. When the stresses in the bonds reach their strength, the bonds break and disappear. The unbonded spheres deform following a Hertzian contact model (Hertz, 1882) and interact by Coulomb friction and rolling friction.

## Particle packing and microstructure approximation

The snow microstructure is approximated by monodisperse spheres, where each sphere represents one snow grain. The initial particle packing is created by ballistic deposition and following removal of random particles. Thus the density of the model snow can be adjusted to a desired value between 250 kg/m$^3$ and 570 kg/m$^3$. By ballistic deposition a random close packing with a volume fraction of 0.62 is obtained. This volume fraction corresponds to a snow density of 570 kg/m$^3$. From this random dense packing random particles are removed until the desired density is reached. Particles can only be removed if the connectivity of the packing is not violated. In a last step neighboring particles are connected by bonds if their distance (from particle center to particle center) is smaller than a certain threshold. This threshold is the parameter “bonding distance”, which controls the coordination number of the structure. Fig. 1 shows a comparison between our model and a real snow microstructure imaged by micro computer tomography. It is obvious that there are large differences between model and reality.

## Parameters

The model parameters can be divided into material parameters and microstructure parameters. The material parameters are based on literature values of ice and summarized in Table 1. All material parameters are fixed parameters for all load cases and snow types and depend only on temperature. All microstructure parameters are variable depending only on the snow type and are summarized in Table 2. We did not consider different temperatures. All experiments were performed at -5 °C.

The determination of the parameters is the most crucial aspect of the model. There are some parameters which are obvious to interpret and to choose and there are other parameters which are difficult to choose and will be fitted by experiments. The latter parameters are indicated by an asterisk (*) in table 1 and table 2.

Obvious parameters are the material parameters density, Young´s modulus and Poisson´s ratio of the particles and strength of the bonds. These parameters correspond directly to literature values of ice. Also the microstructure parameter "density of the packing" is an obvious parameter. Density is a parameter which can easily be measured in real snow. Thus density will be matched between model and real snow.

All other parameters are difficult to choose. Either because they are abstract DE specific parameters which do not have a clear counterpart in real snow, like the rolling friction, or because the parameter can not definitely be obtained from literature values, like the coefficient of ice-ice friction. In the following we discuss these parameters and how they were chosen, in order of appearance in table 1 and table 2.

For the coefficient of ice-ice friction a wide range of values from 0.01 to 0.7 can be found in literature (Schulson and Fortt, 2012)(Yasutome et al., 1999). Therefore we fitted this parameter with an angle of repose experiment. The coefficient of restitution of ice varies from 0 to 0.9 depending on particle velocities (Higa et al., 1996). We have chosen a small value to have more damping and thus a numerically more stable system. The Young´s Modulus of the bonds is adjusted from the Young´s modulus of ice to obtain a better agreement with experimental results. The value is reduced in our model by two orders of magnitude compared to values for ice found in literature (Schulson and Duval, 2009). This can be justified by the fact that snow exhibits a certain creep contribution even at high strain rates. The Young´s modulus of snow determined with high strain rate experiments is up to two orders of magnitude smaller than finite element simulations based on the Young´s modulus of ice (Köchle and Schneebeli, 2014), which were recently shown to correspond to the true elastic modulus of snow (Gerling et al., 2017). The diameter of the spheres is fixed to 1 mm for convenience. For real snow the size of the grains ranges from 0.2 mm to 2 mm. The shape of the snow grains is considered by the rolling friction parameter and will be fitted by the angle of repose experiment. The bond diameter is an important parameter which can be adjusted to consider differently sintered snow. This parameter will be fitted by compression tests of differently sintered snow. The bonding distance which controls the coordination number of the model snow is fixed to 1.1 mm. This is a strong simplification, since the coordination number has a strong impact on the mechanical behavior of granular materials (Gaume et al., 2017) and might differ for different snow types.

The calibration experiments will be explained in the next section.

## Boundary conditions

The boundary conditions are chosen to correspond with the experiments. Dimensions of the snow samples, external forces and external displacements are exactly reproduced in the models. External forces and displacements are applied on the model snow by rigid objects of arbitrary geometry, which can be defined as a 3D geometry in the STL-file format.

**Table 1 Material parameters**

| | Model | Ice property (from literature) |
|---|---|---|
| Density of particle | 917 kg/m$^3$ | 917 kg/m$^3$ |
| Young´s modulus of particle | 9 GPa | 9 GPa |
| Poisson´s ratio of particle | 0.3 | 0.3 |
| Coefficient of friction * | 0.3 | 0.01 – 0.7 |
| Coefficient of restitution * | 0.2 | 0 - 0.9 |
| Young´s modulus of bond * | 90 GPa | 9 GPa |
| Strength of bond | 2 MPa | 1 – 4 MPa |

**Table 2 Microstructure parameters**

| | Model | Real Snow |
|---|---|---|
| Density of packing | 250 – 550 kg/m$^3$ | 30 – 570 kg/m$^3$ |
| Particle diameter * | 1 mm | 0.2 – 2 mm |
| Coefficient of rolling friction * | 0.2 – 0.3 | |
| Bond diameter / particle diameter * | 0.05 – 0.5 | |
| Bonding distance / particle diameter * | 1.1 | |

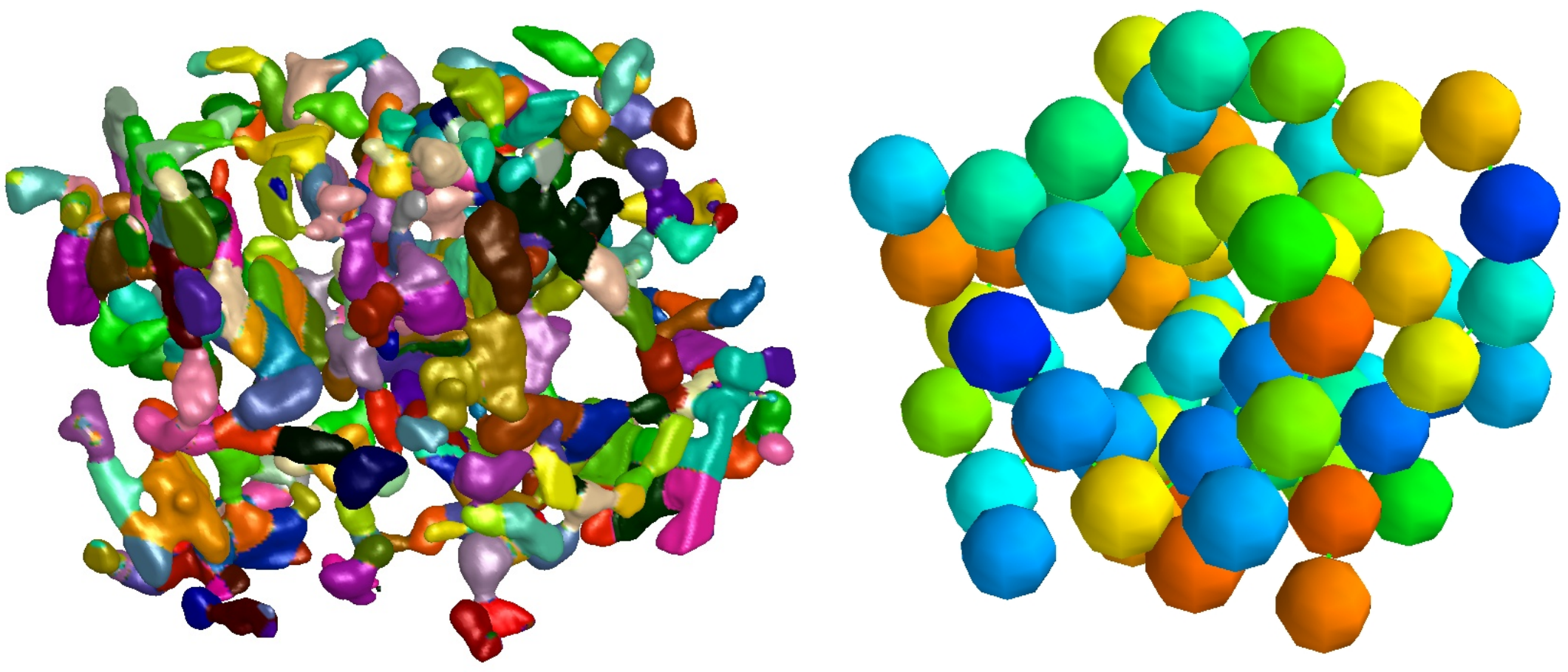

**Fig. 1:** Comparion real snow (left) with model snow (right). The side length of the cubic volumes is 5 mm and the density of both volumes is 270 kg/m$^3$. The real snow sample shows rounded snow and was imaged by computer tomography. Single grains were identified by image processing (Theile and Schneebeli, 2011) and color labeled.

## Experiments

Several experiments were conducted to calibrate and verify the model under different load cases. Emphasis was placed on experiments with large strains where both, the sintered and granular state of snow occurs. During compression and shear deformation large strains can be applied and three phases of deformation can be distinguished: First the elastic deformation of the sintered snow, followed by failure and finally the granular behavior, which is dominated by the interaction of loose snow grains. The failure of the sintered snow will be referred to as fluidization in the following. To analyze the granular behavior separately, we conducted angle of repose experiments.

**Snow types**

Basically three different snow types were used for the different experiments: rounded snow, faceted snow and crushed ice (Fig. 2). Rounded snow develops in several weeks from fresh snow under isothermal conditions by equilibrium metamorphism. Faceted snow grows in a few hours or days under high temperature gradients by kinetic metamorphism. The rounded and faceted snow was collected from an alpine snowpack. The crushed ice was created by freezing tab water and crushing the ice with a commercial ice crusher. All snow types were sieved before the experiments. Only grains which passed the first, large sieve and did not pass the second small sieve were kept for the experiments. Finally the grains were sieved into the sample holder, compressed to a defined density and left for sintering for a defined sintering time. Details about the snow types used for the experiments are summarized in table 3.

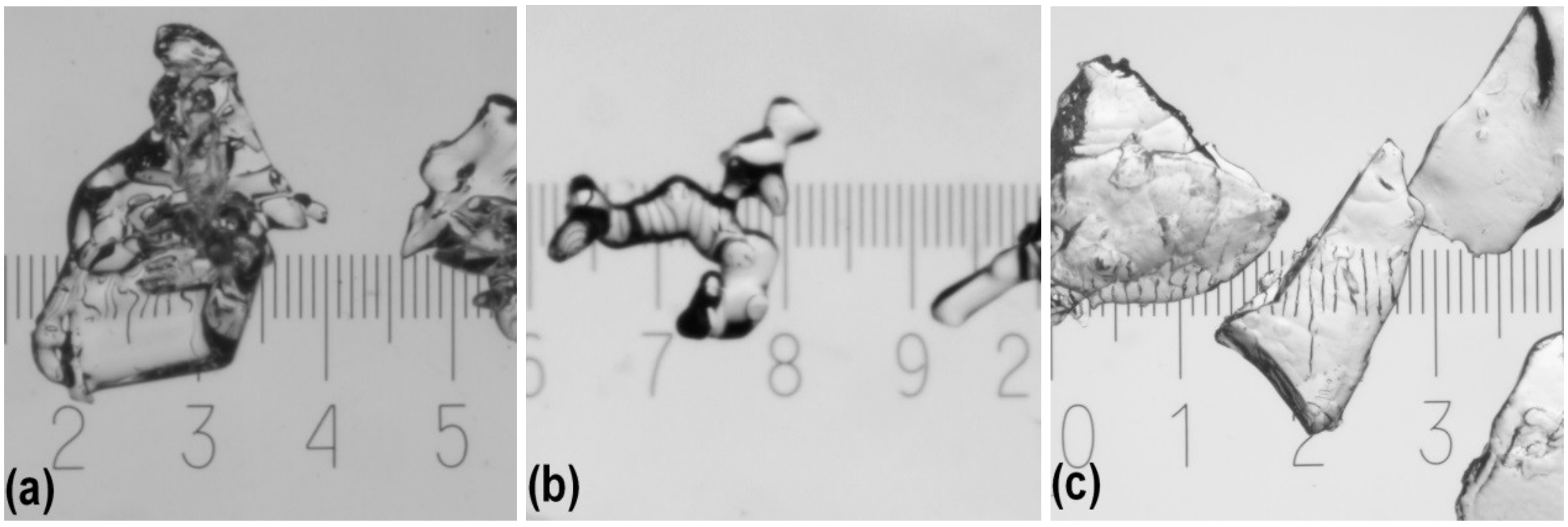

**Fig. 2:** Shape of single snow grains. Image a) shows a faceted grain. Image b) shows rounded grains. And image c) shows grains from crushed ice. The scale in the background is in millimeters. All images have the same scaling.

**Angle of repose**

For the angle of repose experiments snow was sieved onto a cylinder with 50 mm diameter. The angle of the heap, which formed on the cylinder, was measured. The sieve was placed 40 mm above the cylinder. Two different snow types were used: rounded snow and faceted snow (Fig. 2). The temperature was set to -5 °C. Images of the heap from six perspectives were taken to measure the angle all around the heap. The heap which formed is not a perfect cone, it is flattened on the top due to the impacting particles. The angle was measured up to the point where the heap has a constant slope. Fig. 3 a) indicates how the angle was measured.

The goal of this experiment was to calibrate the particle-particle friction and rolling friction parameter for different snow types.

**Table 3 Snow types used for experiments**

| **Experiment** | **Snow type** | **Sieve (mm)** | **Sintering time (hours)** | **Density (kg/m$^3$)** | **Specific surface area (1/mm)** |
|---|---|---|---|---|---|
| Angle of repose | Rounded grains | 0.7 – 1.4 | No sintering | 370 | 11 |
| Angle of repose | Faceted grains | 0.7 – 1.4 | No sintering | 400 | 11.5 |
| Unconfined compression | Rounded grains | 0.0 – 1.4 | 0.03 - 100 | 350 | 25 |
| Unconfined compression | Faceted grains | 0.0 – 1.4 | 0.03 – 100 | 400 | 20 |
| Direct shear | Crushed ice | 0.5 – 0.7 | 3 | 500 | 30 |
| Direct shear | Crushed ice | 0.5 – 0.7 | 25 | 300 | 30 |

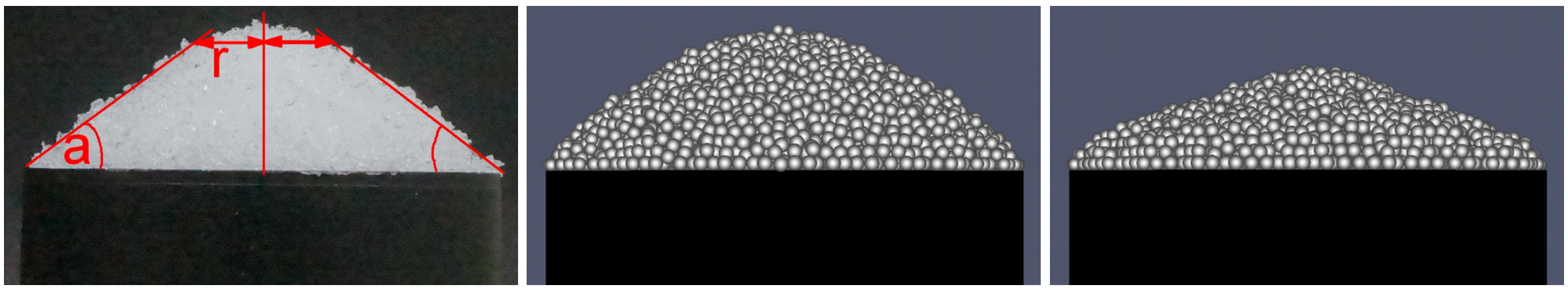

**(a )** **(b)** **(c)**

**Fig. 3:** Comparison of angle of repose experiment and model. Fig. a) and Fig. b) both show an angle of repose of 34°. Fig. a) shows the angle of repose of faceted snow at -5 °C. Fig. b) shows the modeled angle of repose with the model parameters particle-particle friction of 0.3 and rolling friction of 0.2. As a comparison Fig. c) shows the modeled angle of repose of 24° with the model parameters particle-particle friction 0.1 and rolling friction 0.1. The red drawings on Fig. a) indicate how the angle of repose *a* was determined from the images. The angle was measured up to the point where the heap has a constant slope. At this point a certain radius *r* of the cone is reached.

**Unconfined compression**

At a temperature of -5 °C snow was sieved into a cylindrical sample holder of 10 mm height and 40 mm diameter. Two different snow types were used: rounded snow and faceted snow. The samples rested for 1 to 2000 minutes for sintering. Afterwards the surface of the sample was cut using a scraper to obtain a flat, horizontal surface. Subsequently, the side wall of the sample holder was removed. The displacement controlled compression test was executed by pushing a stiff plate with a velocity of 10 mm/s onto the snow sample. Fig. 4 show a snapshot of an unconfined compression experiment. Forces were measured with a precision of 0.5 N and a frequency of 10000 Hz. The highest stress during the first 10% strain is defined as the compressive strength of the sample.

The goal of this experiment is to further calibrate the microstructural parameters of the model, especially how the bond diameter has to be adjusted as a function of sintering time.

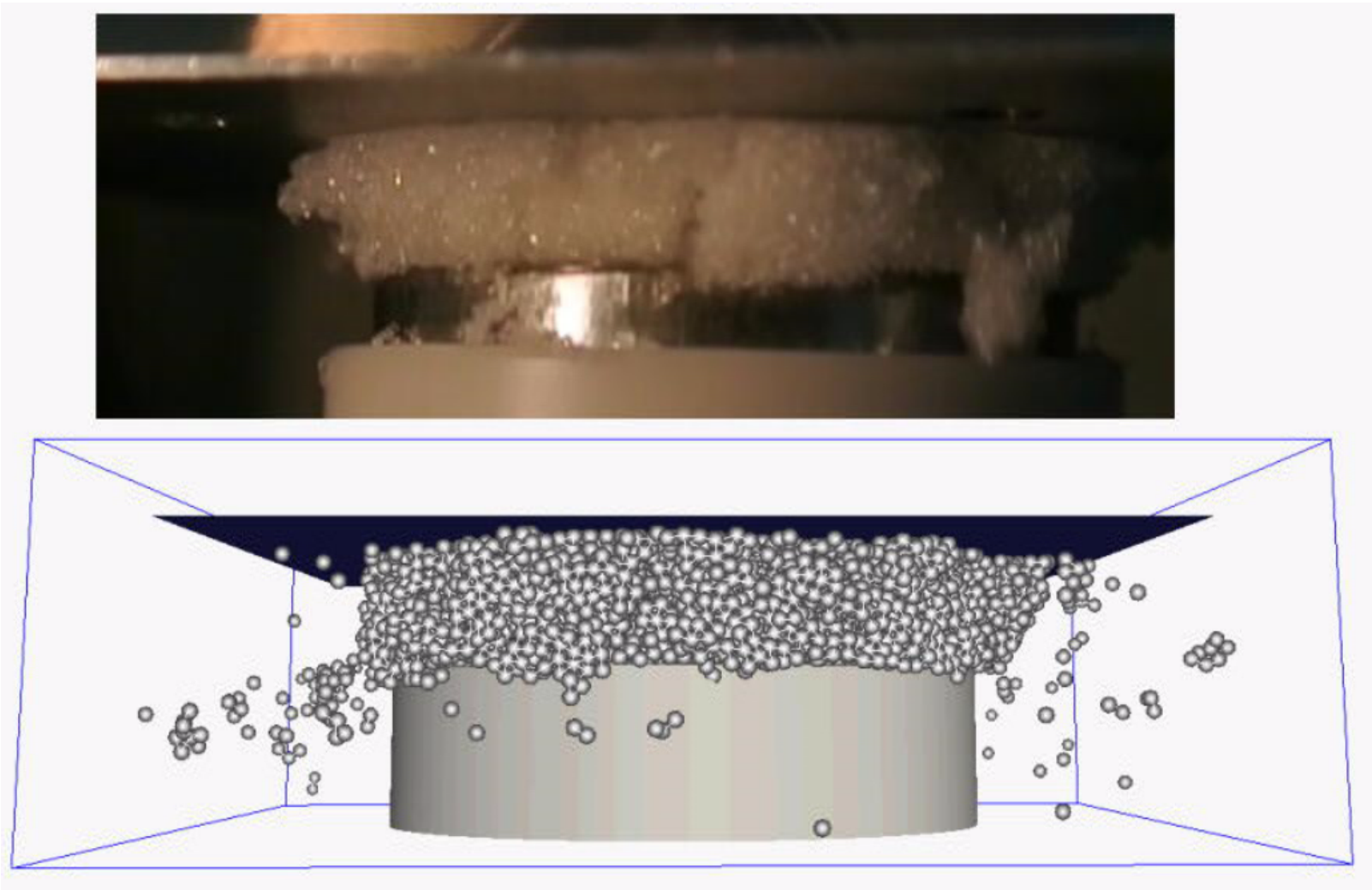

**Fig. 4:** Snapshot of unconfined compression experiment (top) and model (bottom) of a well sintered rounded snow sample after about 30% of strain.

**Direct shear experiment**

At a temperature of −5 °C a rectangular block of snow was placed in a confined shear device. First a vertical compression with a constant velocity of 50 mm/s was applied until a normal stress of 170 kPa was reached, after half a second of pure compression additionally a shear deformation with a constant velocity of 50 mm/s was applied. Snow samples with two different densities were used, “low density” snow with a density of about 300 kg/m$^3$ and “high density” snow with 500 kg/m$^3$. The dimensions of the snow block were 30 mm x 30 mm x 27 mm for the low density snow and 30 mm x 10 mm x 33 mm for the high density snow.

The goal of this experiment is to verify the model with a complex mechanical experiment with high strains where both, the sintered and granular behavior is important.

# Results

### Angle of repose

At a temperature of -5 °C an angle of repose of 36° for the rounded snow and 33.4° for the faceted snow was measured. The experiments were repeated eight times and the standard deviation is about 1°. The measured angles were matched in the simulation with an ice-ice friction of 0.3 and a rolling friction of 0.2 for faceted snow and a rolling friction of 0.3 for the rounded snow. As a comparison, with a rolling friction of 0.1 and a particle-particle friction of 0.1 we obtain a simulated angle of repose of only 24°. Fig. 3 shows a comparison of different heaps which formed during the angle of repose experiment and during the simulation.

### Unconfined compression

The measured stress-strain curves for three unconfined compression experiments with differently sintered rounded snow samples are shown in Fig. 5. The lowest curve corresponds to a sintering time of 2 minutes and reaches a strength of 1 kPa. The highest curve corresponds to a sintering time of 1000 minutes and reaches a strength of 10 kPa. For this curve three different phases of deformation can be distinguished: 1. elastic deformation of the sintered snow; 2. failure; 3. granular behavior. The experimental curves can be matched by adjusting the relative bond size in the model to 0.1, 0.2 and 0.3. From this matching we obtain a relation between sintering time and model bond size. Fig. 6 shows this relation for rounded and faceted snow. This relation follows a power law with an exponent of 0.15 for the rounded snow and 0.2 for the faceted snow. Fig. 7 shows the strength over sintering time for all experiments with rounded snow (red circles) and faceted snow (grey diamonds), as well as the matched simulation results for the two snow types.

Since there are no side walls in a confined compression experiment, we are able to visually observe how the snow fails under pressure. This can also be visualized with the model. Fig. 4 shows a qualitative comparison of the failure behavior in the experiment and in the model. Clearly the cracks in the real and modeled snow sample can be seen.

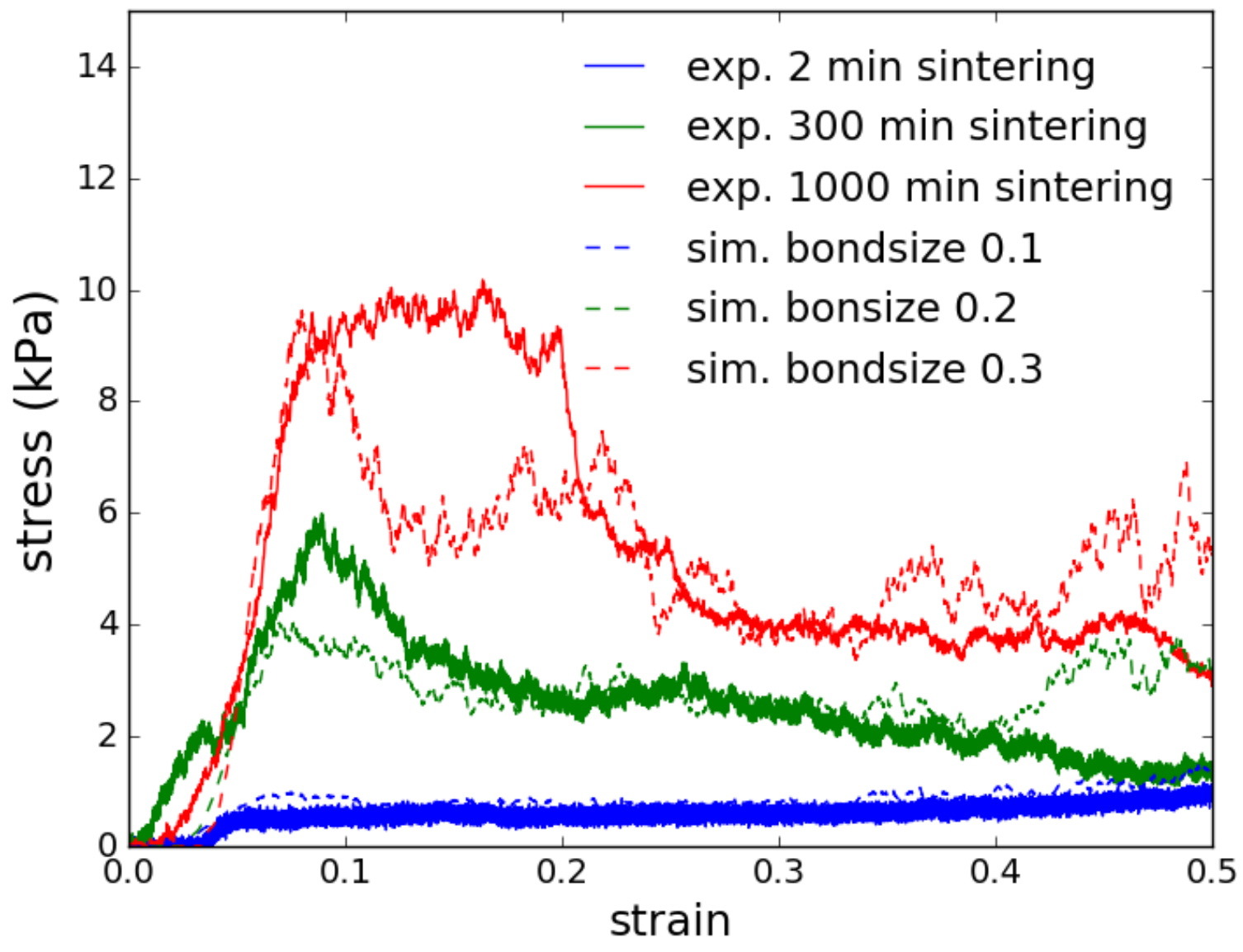

**Fig. 5:** Comparison between simulated and measured stress-strain curves of the unconfined compression. The measured curves differ by sintering time. The simulated curves differ by bond size.

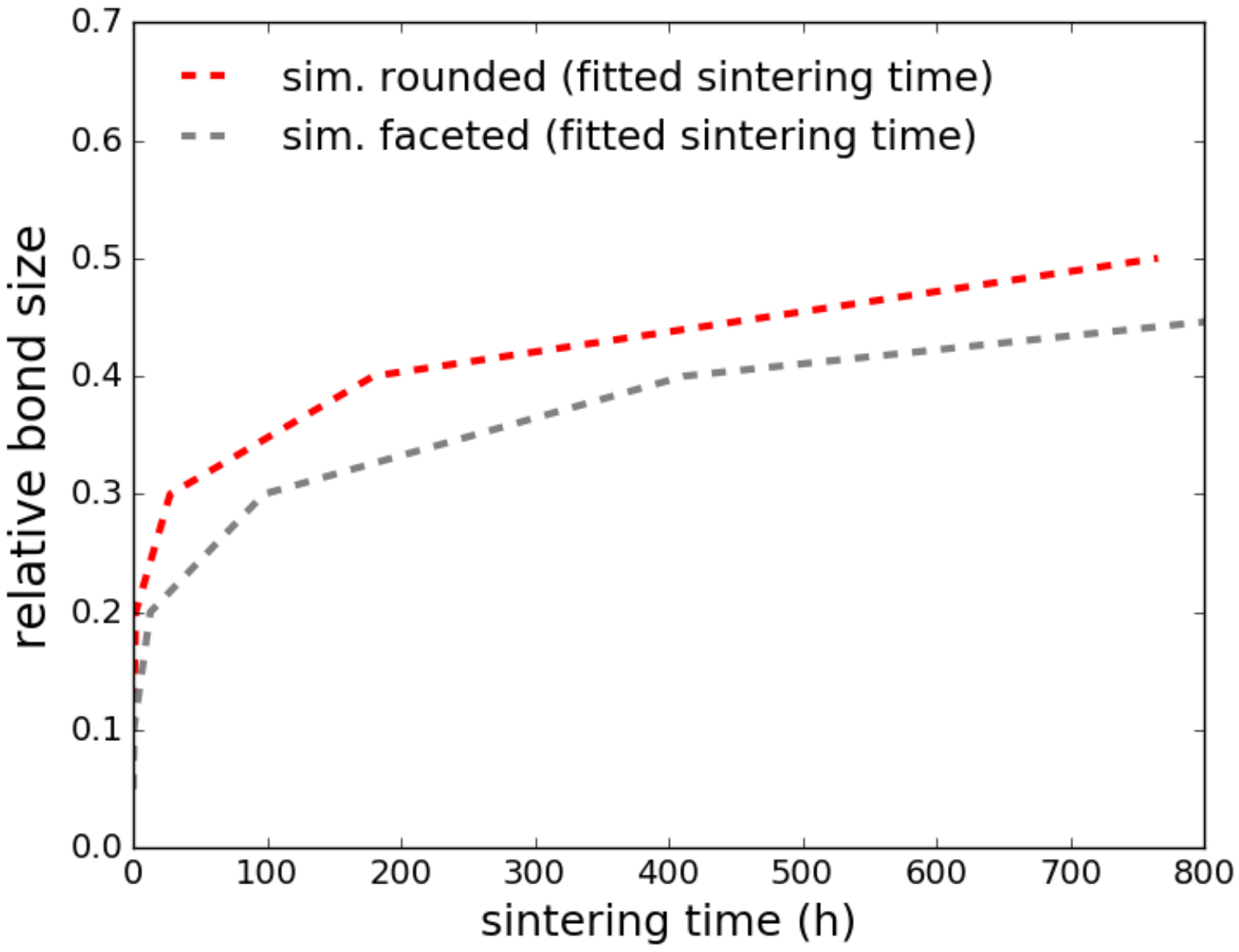

**Fig. 6:** Relation between relative bond size in the model and sintering time. In order to model faceted snow which sintered for 100 hours a relative bond size of 0.3 should be chosen in the model.

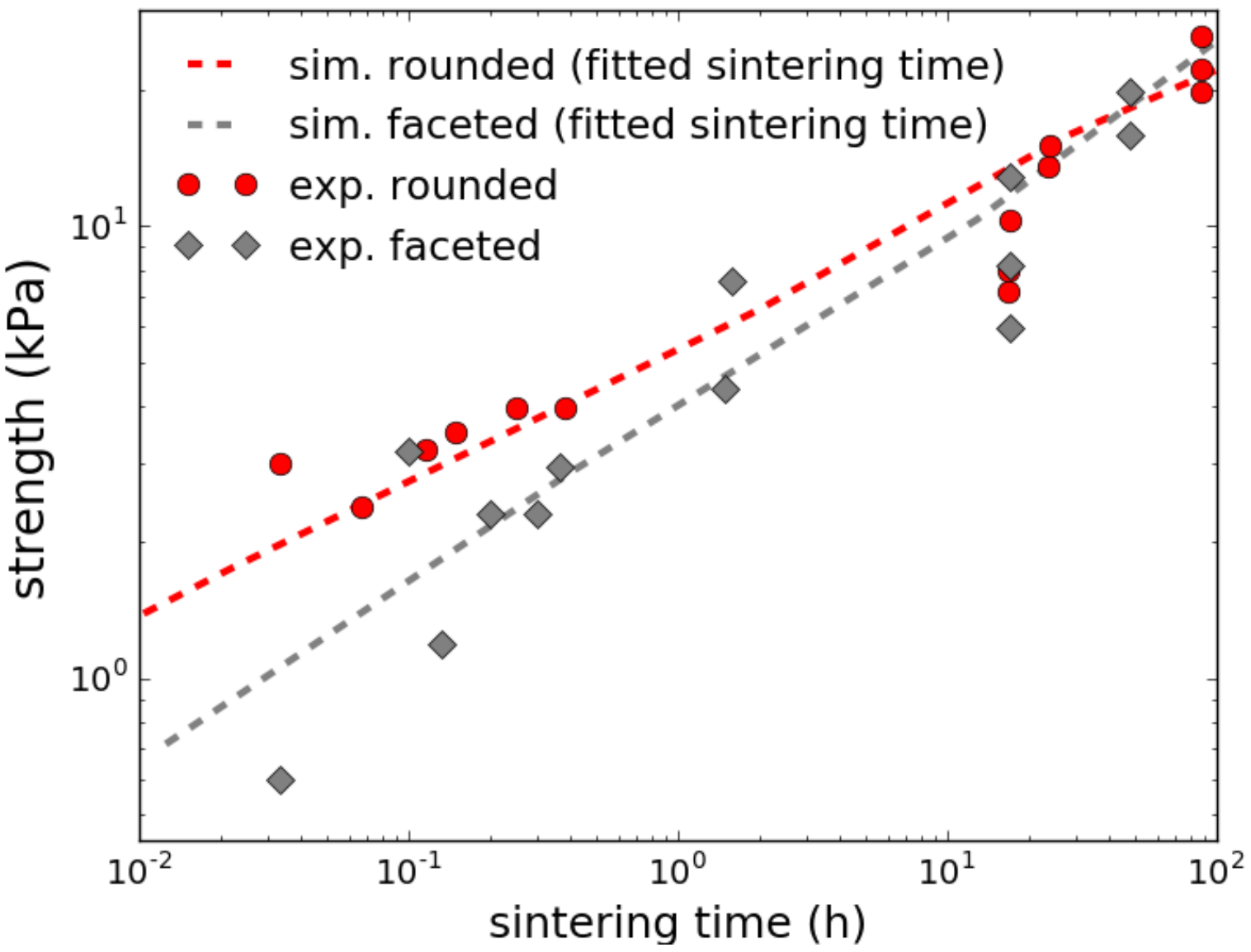

**Fig. 7:** Experimental results of the compressive strength for different sintering time for rounded and faceted snow. The dashed lines show the fitted simulation results.

**Direct shear experiment**

The shear experiment can be divided into two phases: compression and shear deformation (Fig. 8 and 9). First the normal pressure is applied on the snow sample, after half a second the shear deformation starts. The high density snow (500 kg/m$^3$) shows almost no compression at pressures up to 300 kPa. In contrast the low density snow (300 kg/m$^3$) is densified to about 600 kg/m$^3$ at these pressures (Fig. 8). The following shear behavior is also fundamentally different for the low and high density snow. The low density snow shows ideal plastic behavior with pronounced stick-slip during shearing (Fig. 9). In contrast the high density snow shows a brittle fracture behavior with a clear peak stress (shear strength) followed by a drop in shear stress (Fig. 10). The comparisons of the measured and modeled stresses show good agreements. Almost all the characteristics of the compression and shear deformation are captured by the model (Fig. 8-10). Only the stick-slip behavior of the soft snow is not reproduced. The simulations were repeated with a rolling friction of 0.3 (solid black line) and 0.2 (dashed black line).

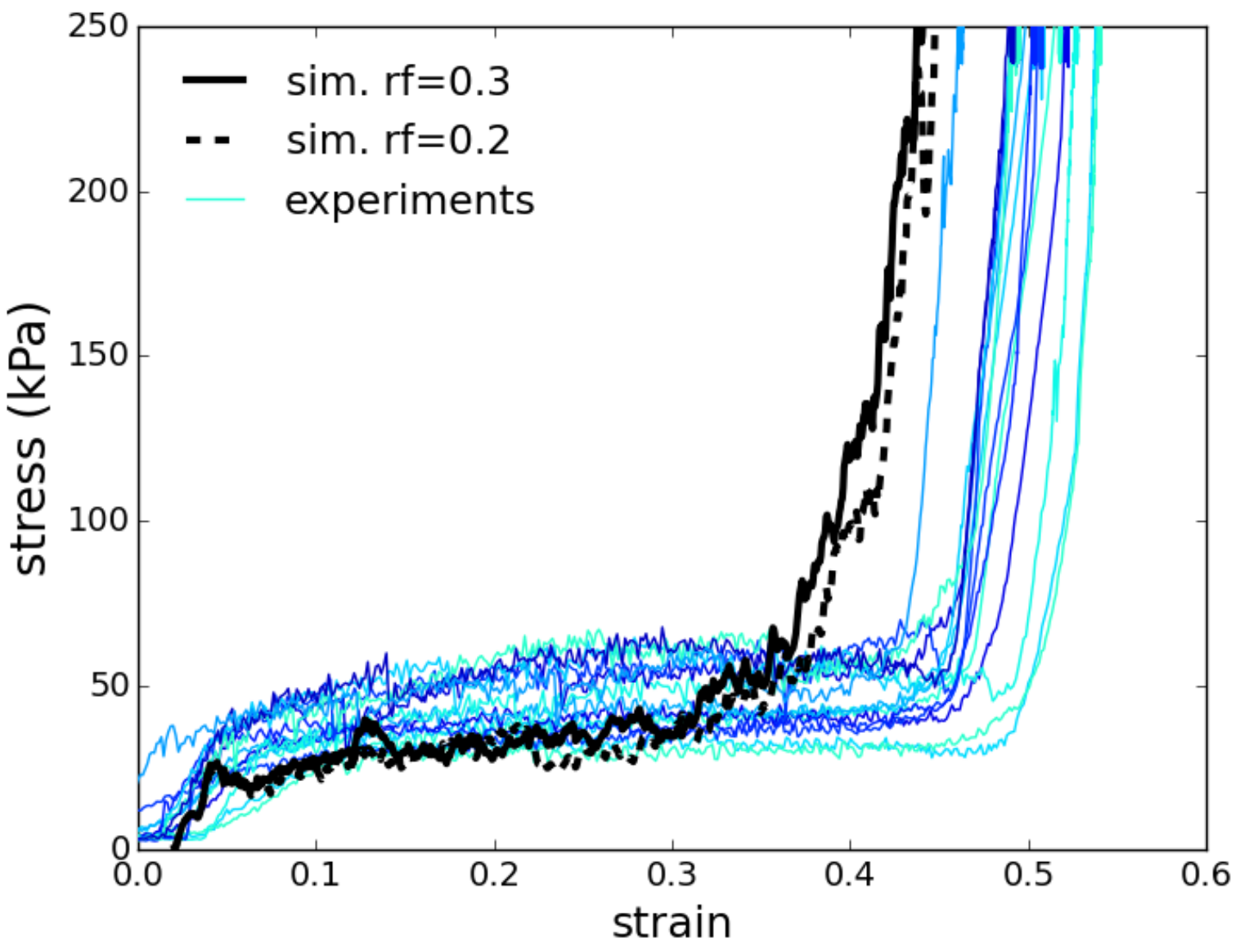

**Fig. 8:** Compression of soft snow. Comparison simulation with experiments. The experiment was repeated 14 times (blue curves). The simulation results with a rolling friction of 0.2 (dashed black line) and 0.3 (solid black line) are shown.

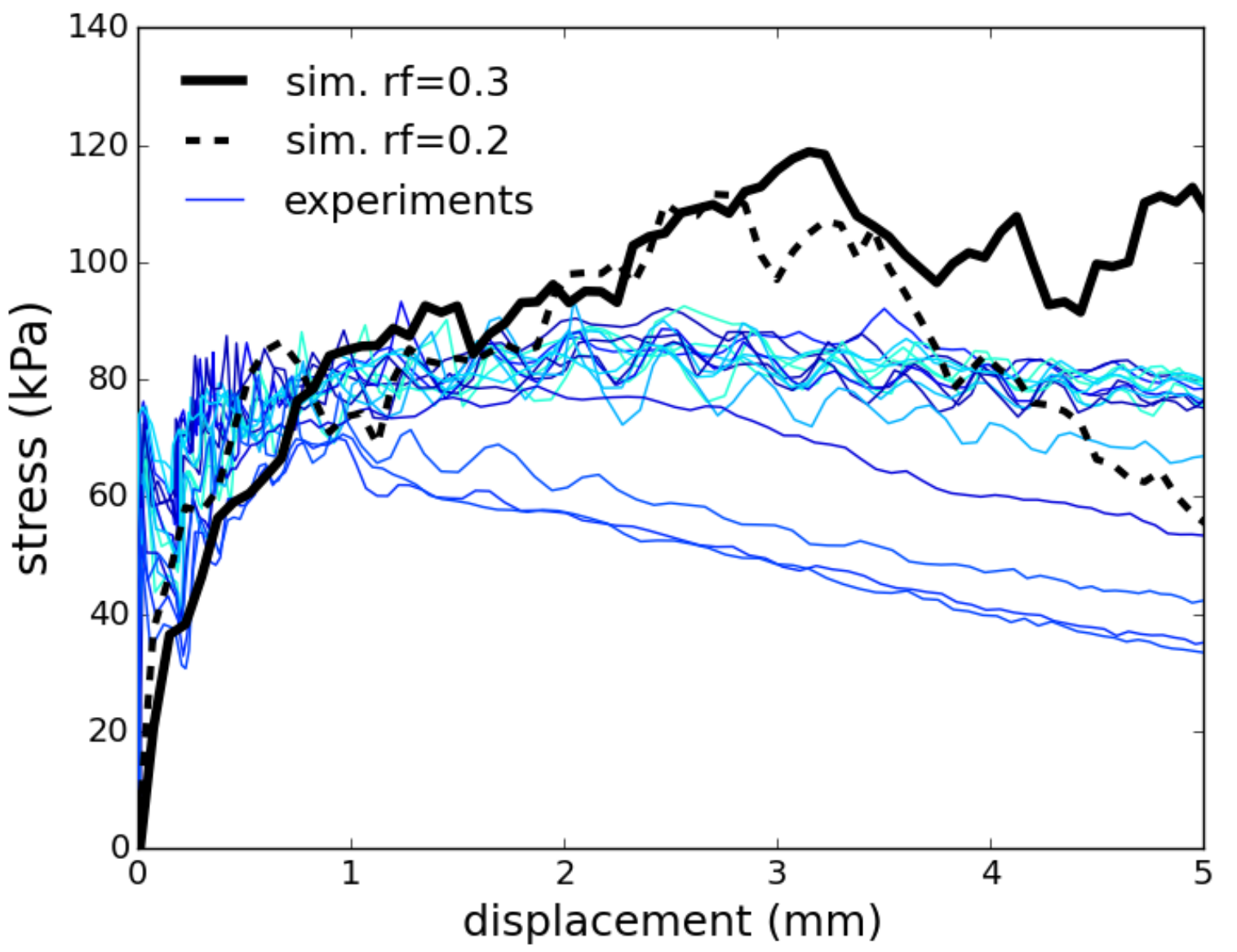

**Fig. 9:** Shearing of soft snow. Comparison simulation with experiments. The experiment was repeated 14 times (blue curves). The simulation results with a rolling friction of 0.2 (dashed black line) and 0.3 (solid black line) are shown.

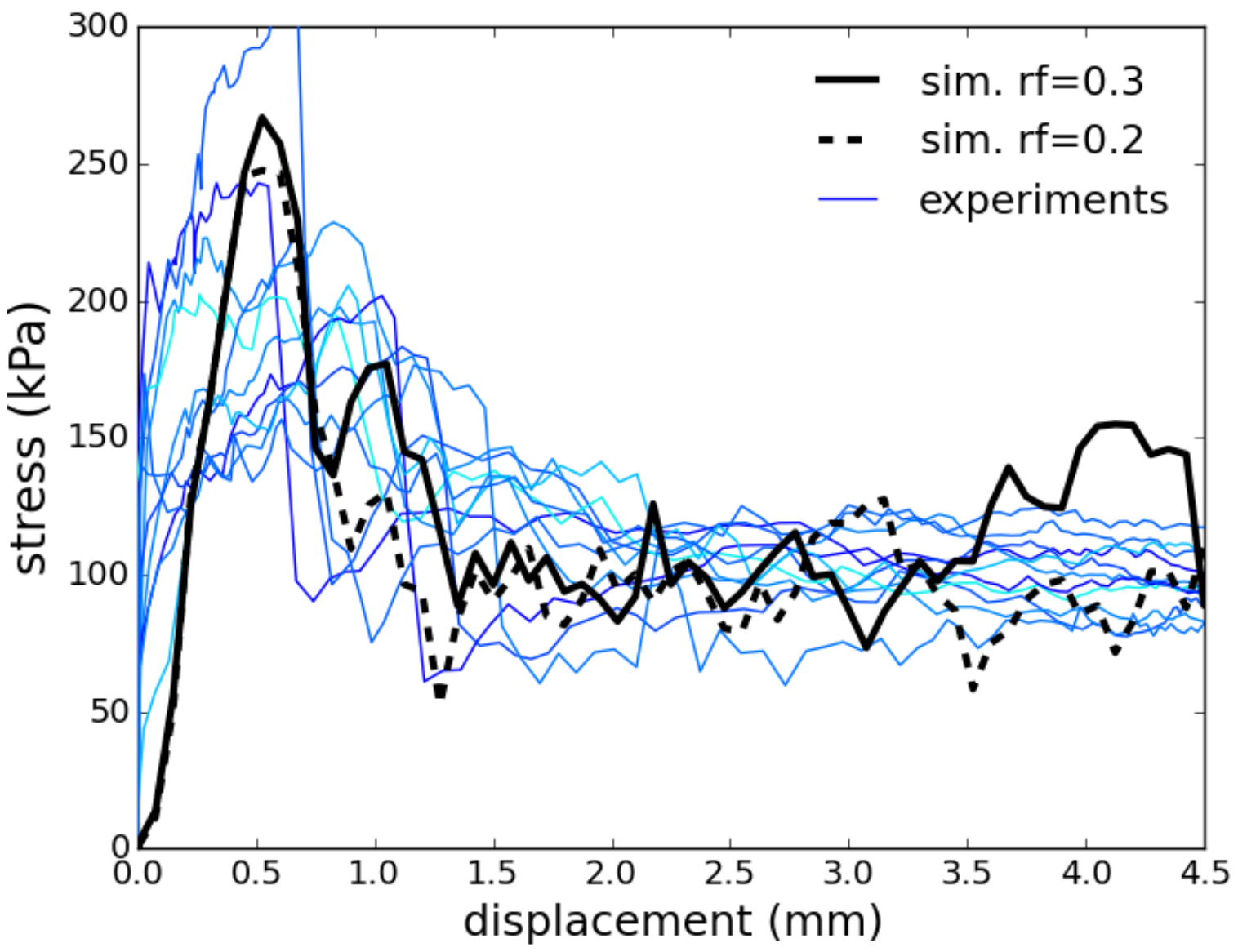

**Fig. 10:** Shearing of hard snow. Comparison simulation with experiments. The experiment was repeated 11 times (blue curves). The simulation results with a rolling friction of 0.2 (dashed black line) and 0.3 (solid black line) are shown.

# Discussion

**Angle of repose**

The angle of repose experiment enables to analyze the granular behavior of snow which results from the interaction of loose ice grains. The interaction of the loose ice particles is controlled mainly by ice-ice friction, the particle shape and sintering. The goal of this experiment is to calibrate the model parameters rolling friction and particle-particle friction for the temperature of -5 °C. The rolling friction parameter is a structural parameter and considers the non-sphericity of the snow grains. The particle-particle friction is a material parameter and corresponds to ice-ice friction. It is very difficult to determine the ice-ice friction from literature. Values between 0.01 and 0.7 (Schulson and Fortt, 2012) can be found. The fitted value of 0.3 seems reasonable.

The rolling friction is a simplification to consider the shape of the snow grains in the model. A physically more correct implementation would approximate the shape of real snow grains. In DE modelling “clumps of spheres“ are often used to approximate the shape of the particles. However, this approach is more difficult to implement and computationally more expensive. Therefore we used the simple rolling friction to consider the non-sphericity of the snow grains.

Sintering is not included in our model. However, it is known that sintering can have a big impact on the mechanical behavior of snow at low strain rates (Reiweger et al., 2009). Also at high strain rates, sintering might have an impact on the mechanical behavior. Fast sintering of ice on the sub-second timescale was described by Szabo and Schneebeli (Szabo and Schneebeli, 2007). Fast sintering is temperature dependent and occurs at temperatures between -15 °C and 0 °C. Fast sintering and its temperature dependence is the reason why you can form snowballs only at warm temperatures close to 0 °C. As sintering is not included explicitly in our model, it will be included in the fitted friction parameter. However, without any time or velocity-dependence.

With this calibration we can estimate the model parameters rolling friction and particle-particle friction. This is useful since the estimation of the two parameters from particle shape and from literature is not possible.

**Unconfined compression**

The unconfined compression experiment was performed to calibrate the bond size in the model. This parameter is the most crucial parameter in the model. The bond size has a small influence on the sintered snow behavior: the thicker the bonds the stiffer the snow. But more importantly the bond size together with the bond strength determines under which stresses the bonds break and how the sintered snow transforms to granular snow, both in the model and in real snow. Since we fixed the bond strength in the model to the strength of ice, we only have to calibrate the bond size. Bond size is a microstructure parameter. This means that different snow types have different bond sizes. It would be elegant to measure the bond size in real snow and feed the model with these values. However, this approach seems not applicable for our model. First it is very difficult to measure the bond size in snow, and second it is unlikely that these bond sizes are valid in our model due to other simplifications of the microstructure. Therefore we just used this parameter as a fitting parameter and used the unconfined compression experiment for calibration. In this experiment we vary the bond size without changing any other characteristics of the snow, like density and grain shape. With increasing sintering time the bonds grow thicker and the strength increases. Not only the strength increases, also the failure behavior changes. For a sintering time of two minutes the snow deforms ideal plastically (blue curve in Fig. 5). With longer sintering time the snow becomes more brittle (green and red curve in Fig. 5). Not only the increasing strength of the snow can be reproduced by increasing the bond size in the model, but also the different failure behavior. This shows how well this model is able to reproduce snow behavior. However, the sintered behavior is not reproduced well for this experiment. For the red and green curves in Fig. 5 the modeled curves show a much stiffer behavior than the experimental curves. Most likely the reason for this difference is that in the experiments the surface of the snow samples is not perfectly parallel to the plate which is pressed on the snow. If the angular deviation is only 1° this will result in an additional strain of 7% until the two surfaces are in full contact. Also the steepening shape of the red curve confirms this explanation.

Hobbs and Mason (Hobbs and Mason, 1964) have shown experimentally and theoretically that bond size increases with sintering time according to a power law with exponent 0.2. The dominant mechanism is vapor transport. The relation between bond size and sintering time shown in Fig. 6 also follows a power law with an exponent of about 0.2 for the faceted snow and 0.15 for the rounded snow. This indicates that the fitted bond sizes behave realistically. Several studies about the sintering of snow have determined a connection between some measured mechanical property and sintering time. Jellinek (Jellinek, 1959) found an exponent of 0.21 for the compressive strength. Van Herwijnen and Miller (Van Herwijnen and Miller, 2013) found an exponent of 0.18 for the penetration resistance and argued that this is in good agreement with sinter theory of ice. However, linking the mechanical property linearly to bond size is doubtful. Using our model we get a direct connection between bond size and sintering time, which allows a comparison to sinter theory and thus a further confirmation of the model.

**Direct shear experiment**

The aim of the direct shear experiment is to further validate and challenge the model with a different load case and with different snow densities. The challenge is to model this complex process with compression and shear deformation with large strains, including the transformation from sintered to granular snow. The two snow types investigated, low- and high-density snow, show fundamentally different behavior. The low-density snow is fluidized already when the compressive stress is applied, the following shear deformation of the granular snow resembles ideal plastic behavior. In contrast the high-density snow hardly deforms when the compressive stress is applied. The strength is higher than the applied stresses. Therefore the fluidization takes place during the shear deformation, showing the typical peak stress at failure followed by granular behavior. The granular shear behavior, which is basically snow-snow friction, is similar for both snow types with a shear stress of around 90 kPa at a normal stress of 170 kPa. This corresponds to a snow-snow friction of 0.53 or a friction angle of 28°. This agrees well with the avalanche rule of thumb saying that a minimum slope of 30° is required for avalanche release.

Interesting is also the compression of the low-density snow. Up to about 50% of strain the snow is densified with an almost constant stress. After this point the stresses suddenly increase significantly. The reason for this behavior is that the low density snow reaches a critical density of 600 kg/m$^3$ after 50% of strain. This density corresponds to the density of a random dense packing. Up to this density snow can be densified by the rearrangement of snow grains. After this point the grains themselves have to be deformed, resulting in a much higher compression resistance. The model also shows this behavior, but the critical density is lower with 500 kg/m$^3$. A possible explanation for this difference is that all particles have the same size in the model, while in real snow particle sizes are distributed and can therefore reach a higher critical density.

The model performs well in reproducing the experimental results with all its characteristics without further fitting of the model parameters. Even though the model parameters were obtained under different load cases. This shows that the most important mechanical processes are included in the model and that the model seems promising to simulate arbitrary high strain rate processes for a wide range of snow types with different densities. However, not all characteristics of the measured results are reproduced by the model. The stick-slip behavior of the low-density snow during shearing is not reproduced by the model. The reason for this stick-slip behavior might be periodic sintering and breaking of snow grains or periodic jamming of irregular shaped snow grains. Both effects are not included in the model. As you would expect, the rolling friction parameter has no impact on the sintered snow behavior but on the granular behavior (dashed and solid black lines in Fig. 8-10). The simulations with a rolling friction of 0.2 correspond slightly better to the experimental results than the simulations with a rolling friction of 0.3.

**Simulation volumes and computing times**

For all experiments the volume of the snow samples was in the order of magnitude of 10 $cm^3$. These volumes correspond to about 10,000 particles with 1 mm diameter in the simulations. The computing times ranged from 20 minutes to 240 minutes on one processor depending on the number of time steps. For the angle of repose simulation, which is without bonds, 14 million time steps are calculated per second for one particle. For the simulations with bonds about 1 million time steps are calculated per second for one particle. The time step was set to $10^{-7}$ seconds for all simulations.

To check the potential of the model for larger volumes a simulation with one million particles and 0.4 million time steps was calculated on a cluster on 80 cores in 1320 minutes. The same simulation was 3.7 times faster on four times more cores, showing the potential of the parallelization for large problems.

Table 4 Computing times

| Model | # of particles | # of bonds | # of time steps (in million) | # of processors | Computing time (in minutes) |
|---|---|---|---|---|---|
| Angle of repose | 10000 | 0 | 20 | 1 | 240 |
| Unconfined compression | 12000 | 35000 | 1.9 | 1 | 160 |
| Confined | 9000 | 20000 | 1.4 | 1 | 200 |
| compression | | | | | |
| Shear deformation (low density snow) | 18000 | 41000 | 0.1 | 1 | 30 |
| Shear deformation (high density snow) | 8000 | 27000 | 0.1 | 1 | 20 |
| Large test simulation | 1000000 | 1500000 | 0.4 | 320 (80) | 360 (1320) |

## Conclusions

We have developed and verified a DE model for high strain rate deformations of snow. The transformation from sintered to granular snow is the key process for large and fast deformations of snow. The model includes this mechanism. Furthermore the model considers different characteristics of the snow microstructure, like density or grain shape. Including relevant micromechanical processes and characteristics of different snow types is the key to create a snow model as general as possible. In the presented model the complex microstructure of snow is approximated by a "simple" sphere system. Nevertheless, the model performs well in reproducing different experiments with different snow types and different load cases with one set of parameters. This versatility is the strongest point of the model. Another strong point of the model is the free and easy availability of the model. All presented simulations can be found in the supplementary materials. Weaknesses of the model are the missing study of temperature dependence, the missing implementation of sintering and the simplified material model without creep deformation. However, these simplifications are a tradeoff between accuracy and calculation time, they keep the model simple and reduce the number of parameters. Due to this simplicity and the fast and parallized liggghts-code, simulations with up to two million particles can be solved in reasonable times. This enables the application of this model to different snow engineering problems, like mobility on snow or snow removal equipment, in future work.